# Mode-hop-free tuning over 135 GHz of external cavity diode lasers without anti-reflection coating

Sourav Dutta,[1,*] D. S. Elliott,[1,2,‡] and Yong P. Chen[1,2,†]

[1]*Department of Physics, Purdue University, West Lafayette, IN 47907, USA*
[2]*School of Electrical and Computer Engineering, Purdue University, West Lafayette, IN 47907, USA*
[*]*Corresponding author: sourav.dutta.mr@gmail.com*
[‡]*elliottd@purdue.edu*
[†]*yongchen@purdue.edu*

**ABSTRACT:** We report an external cavity diode laser (ECDL), using a diode whose front facet is not anti-reflection (AR) coated, that has a mode-hop-free (MHF) tuning range greater than 135 GHz. We achieved this using a short external cavity and by simultaneously tuning the internal and external modes of the laser. We find that the precise location of the pivot point of the grating in our laser is less critical than commonly believed. The general applicability of the method, combined with the compact portable mechanical and electronic design, makes it well suited for both research and industrial applications.

## 1. INTRODUCTION

Single-mode, narrow band, continuous-wave (cw) lasers exhibiting wide and continuous wavelength tunability are of considerable interest for a variety of applications including high resolution spectroscopy [1], cooling and trapping [2], as well as sensing [3]. Semiconductor lasers are often used because their lasing wavelength can be easily tuned by changing either the injection current or the chip temperature. Distributed Bragg reflector (DBR) and distributed feedback (DFB) lasers provide large tuning range along with narrow linewidth and excellent power and frequency stability. However, DBR/DFB lasers are relatively expensive and available only at a few selected wavelengths. Fabry-Perot (FP) diode lasers, on the other hand, are cheap and available in a variety of wavelengths, but their tunability can be limited. The performance of a FP diode laser can be significantly improved by placing it within an external cavity employing a diffraction grating, either in a Littrow [4,5] or a Littman [6] configuration. This arrangement substantially reduces the linewidth of the output beam, and allows coarse wavelength tuning; in the case of the Littrow configuration, through rotation of the grating.

Several groups have addressed the issue of increasing the mode-hop-free (MHF) tuning range of single-mode Littrow-type lasers. It has been proposed [7,8] that MHF tuning range can be enhanced by rotating the grating about an optimal pivot point, located precisely at the intersection of the grating plane and plane of the diode's rear facet (Fig. 1). Several works [9,10] discussed the sensitivity of the behavior of the laser on the precise location of this pivot point, while other works [11-13] focused on tuning the internal and external cavity lengths in the proper proportion. The maximum MHF tuning range reported for ECDLs without electronic feedback [14] or elaborate schemes involving acousto-optic modulators [15] is 110 GHz [10], achieved with AR coated diodes. For non-AR coated diodes, the maximum MHF tuning range reported is 90 GHz [11].

In this article, we report a simple, single-mode, narrow band, cw Littrow-type ECDL with a 137 GHz MHF tuning range using an "off the shelf" laser diode without AR coating. This, to the best of our knowledge, is the largest MHF tuning range among ECDLs using diodes with or without AR coating. Our observations lead us to conclude that the short external cavity length, in concert with careful scaling of the variations of the internal and external cavity lengths, are critical to achieving large MHF tuning range, while the precise location of the pivot point of the grating is less important.

In the next section, we review the principle behind continuous MHF tuning. This is followed by the details of the mechanical design of our ECDL and the method we use to maximize the MHF tuning range. Measurements of the tuning range, including applications in spectroscopy, are then presented, followed by the conclusion.

## 2. PRINCIPLE

The key factors that determine the frequency of the active mode of an ECDL have been discussed in a number of previous works; see for example [5,10,12]. In short, an ECDL tends to lase at the frequency with the greatest gain and lowest loss. The free running laser (FRL) has a broad gain profile (typical width ~ 10 nm). The internal cavity modes of the laser diode are typically spaced by around 100 GHz and the internal mode that experiences maximum gain determines the emission wavelength of the FRL. When the FRL is placed in an external cavity, the internal and external cavities are optically coupled. Frequencies of coupled cavities depend on the optical length of the two individual cavities and the reflectivity of the surfaces in a complex way [16]. In a simpler approximation, one can assume that the wavelength of the ECDL is determined by the external cavity mode positioned most closely to the center of the internal cavity mode.

For a Littrow-type ECDL, the wavelength $\lambda$ that the grating feeds back to the diode chip is determined by the equation:

$$2a \sin \theta = \lambda \quad (1)$$

where $\theta$ is the angle of incidence (see Fig. 1) and $a$ is the line spacing of the grating. The angular width $\Delta\theta_g$ of the beam diffracted from the grating can be shown to be $\Delta\theta_g = \lambda/(N a \cos\theta)$, where $N$ is the number of grating lines illuminated by the laser beam [17]. Using representative parameters for our ECDL, where we have a grating with $a^{-1} = 1800$ lines/mm, a beam size of ~ 3 mm (thus $N$ is about 5400), $\lambda = 780$ nm and $\theta \sim 44.6°$, we find that $\Delta\theta_g$ is ~ 0.37 $mrad$. From Eq. (1), the grating rotation $\Delta\theta_{max}$ required for a scan of $\Delta\upsilon_{max}$ is found to be: $\Delta\theta_{max} = \Delta\upsilon_{max} \lambda^2 / (2ca \cos\theta)$, which for our 780 nm ECDL with $\Delta\upsilon_{max} \sim 140$ GHz yields $\Delta\theta_{max} \sim 0.36$ $mrad$. Since $\Delta\theta_{max}$ is comparable to $\Delta\theta_g$ (in fact $\Delta\theta_{max}/\Delta\theta_g \sim 1$ and is independent of $\theta$), the precision of the rotation angle (and hence the pivot point of the grating) is not very critical for our range of tuning.

Another key factor that determines the frequency of the ECDL is the length $L_e$ (~ 15 mm for our ECDLs) of the external cavity. The resonant external cavity mode is given by the equation:

$$m \lambda / 2 = L_e \quad (2)$$

where $m$ is an integer. The transmission peaks of the external cavity are spaced by the cavity free spectral range (FSR) of ~ 10 GHz, much closer than those of the internal cavity.

To obtain a broad continuous MHF tuning range, the frequency supported by all the above factors must tune synchronously. As is commonly done [10-13], we tune the gain profile of the diode and the internal cavity mode frequency by varying the injection current $I_{LD}$. The amount by which the frequency changes with a change in the injection current is quantified through a parameter $\beta = \Delta\upsilon / \Delta I_{LD}$, and can easily be determined experimentally for a free running diode. The external mode frequency is tuned by varying the voltage applied to a piezoelectric transducer (PZT) that controls the grating position [10,13]. The variation of the external mode frequency is characterized by a parameter $\alpha = \Delta\upsilon / \Delta V_{PZT}$, which we determine experimentally, as described in the following section.

To summarize, the wavelength of the ECDL at a fixed temperature is determined by the injection current $I_{LD}$, the grating angle $\theta$ and the length $L_e$ of the external cavity. To obtain MHF tuning, the shift ($\Delta\upsilon = \beta \Delta I_{LD}$) of the internal mode frequency caused by the change in $I_{LD}$ must be equal to the frequency shift ($\Delta\upsilon = \alpha \Delta V_{PZT}$) caused by the change in cavity length $L_e$ and, to a lesser degree, the grating angle $\theta$.

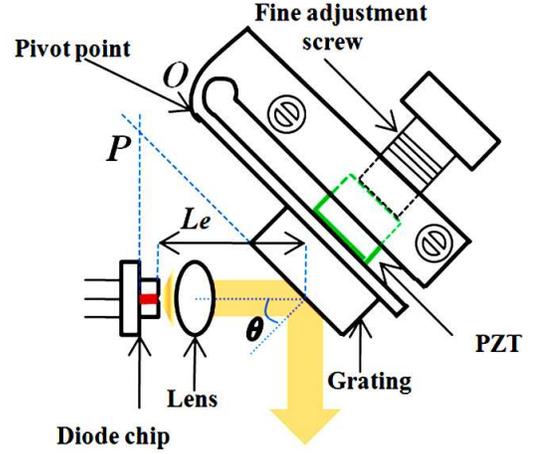

Fig. 1. Schematic of the mechanical setup of the ECDLs. The pivot point $O$ of the grating is close to, but not precisely at, the optimal pivot point $P$. $L_e$ is measured along the beam axis from the front facet of the diode chip to the grating.

## 3. EXPERIMENTAL DESIGN

We use commercial single-mode diode lasers without an AR coating in all our ECDLs. We have used many different diodes (Sanyo DL7140-201S, Hitachi HL7851G, Arima Lasers ADL-78901TL and Lasermate LD-808-100A) and all of them have scanned more than 120 GHz without mode hops. The lengths of the diode chips are less than 1 mm. The physical assembly of our design is similar to that described in Ref. 5 with some modifications. The beam is collimated with a lens (Thorlabs LT230P-B) and is directed onto a holographic diffraction grating with 1800 lines/mm (Thorlabs

GH13-18V). The homemade grating mount has a lever arm that rotates about the pivot point $O$ (Fig. 1). A piezoelectric actuator (Thorlabs AE0505D08F) is clamped between the lever arm and the fine-pitched adjustment screw that provides the horizontal adjustment by rotating the grating (in the plane of the figure). The vertical, out of the plane, adjustment is provided by another fine-pitched screw attached to the base (not shown in the figure). The whole base is temperature stabilized with a thermoelectric cooler. The injection current is supplied by a commercial laser diode driver (Thorlabs LDC 205C) and the temperature is controlled by a temperature controller (Thorlabs TED200C). The manufacturer and part numbers are mentioned only for the completeness of the description and not for endorsement.

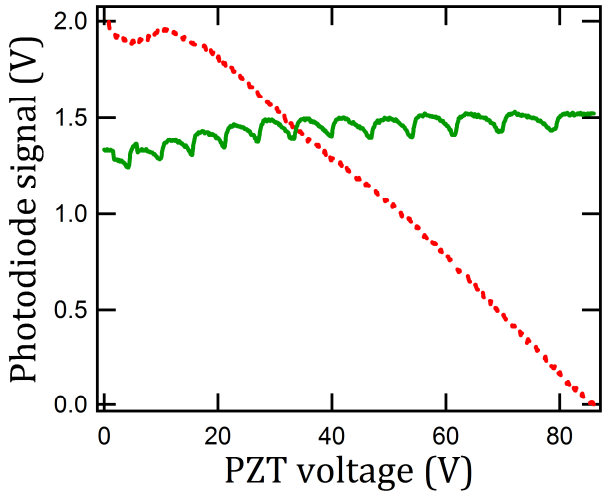

Fig. 2. Solid line (green): Power variation (measured with a photodiode) of an 808 nm ECDL on scanning the PZT with a linear ramp of 86V at a fixed value of injection current, resulting in multiple mode-hops. Dotted line (red): Power variation during a MHF scan of 121 GHz obtained with simultaneously varying the injection current and the PZT voltage.

In order to maximize the MHF tuning range of the ECDL, we must first determine $\beta = \Delta v / \Delta I_{LD}$ for our free-running diodes, and $\alpha = \Delta v / \Delta V_{PZT}$. For the former, we vary the injection current $I_{LD}$ in small steps and record the frequency $v$ on a wavelength-meter (Bristol Instruments 621A, resolution ~60MHz). The measured values of $\beta$ are within the range $-1.4$ GHz/mA to $-2.4$ GHz/mA for each of the different diodes we tested, and we estimate the measurement accuracy to be 0.1 GHz/mA. $\beta$ varies as a function of $I_{LD}$. To determine $\alpha$, we mount the diode in the external cavity, and measure the output power of the laser as a function of $V_{PZT}$. We show an example of this measurement in Fig. 2. Changing the external cavity length $L_e$ by varying $V_{PZT}$, while holding the injection current fixed, tunes the external cavity mode frequency across the peak of the internal cavity mode. The output power is maximized when the external and internal mode frequencies are the same, but decreases as the mismatch increases. Mode hops occur when the shift in the external mode frequency exceeds half the FSR of the external cavity. At this point, the adjacent external mode has moved closer to the peak of the internal mode, and the laser oscillation hops to this new mode. In the green curve (solid line) of Fig. 2, these mode hops are seen at each of the minima, which are spaced by one free spectral range (FSR) of approximately 10 GHz. The corresponding change in $V_{PZT}$ yields $\alpha \sim 1.5$ GHz/V near the center of this figure. The voltage spacing between peaks varies with $V_{PZT}$ due to the nonlinear behavior of the PZT. The 12 mode hops recorded for the scan of 86 V, together with the 10 GHz FSR of the ECDL, suggests that a maximum scan range of $\Delta v_{max} \sim$ 120 GHz (= 12 x FSR) should be possible for this example. Indeed, concurrent scanning of $I_{LD}$ and $V_{PZT}$ yields the red curve (dotted line) in Fig. 2. The absence of peaks and valleys in this data is characteristic MHF tuning of the output frequency. This is confirmed by the smooth variation of the output frequency of the laser as measured with an optical wavelength meter. The nonlinear behavior of the output power when $V_{PZT}$ is less than 15 V may be due to the variation of $\alpha$ and $\beta$ over the long frequency scan and coupled cavity effects of the internal and external cavities. We have used this procedure to obtain MHF scan of over 120 GHz for all the diodes that we used, and up to 137 GHz for one of the diodes as discussed in the next section.

## 4. EXPERIMENTAL RESULTS

In this section, we present our results for one of our ECDLs with a 780 nm laser diode from Arima Lasers (ADL-78901TL). The design for this ECDL is based on Fig. 1. The output beam is elliptical in shape, and remains collimated without noticeable divergence up to 2 meters. We use a Fabry-Perot interferometer of moderate finesse to place an upper bound on the laser linewidth of ~2 MHz. We determined the average value of $\beta$ to be $-1.7$ GHz/mA ($\beta$ varies between $-1.4$ GHz/mA and $-1.9$ GHz/mA as a function of $I_{LD}$.) $\alpha$ is determined to be 1.5 GHz/V, using a measurement similar to that shown in Fig. 2. The optimal scaling between the injection current and the PZT voltage is then estimated to be $\Delta I_{LD} / \Delta V_{PZT} = \alpha / \beta \approx -0.88$ mA/V. We optimize this value empirically by determining the MHF tuning range at a variety of different values, and find the optimal value to be $\Delta I_{LD} / \Delta V_{PZT} = -0.93$ mA/V. We show the frequency scan of the ECDL, measured using a wavelength meter, in Fig. 3 as the solid (red) curve. The dotted (green) curve shows the power variation of the ECDL. The longest frequency scan without mode hops was 137 GHz, accomplished with a 90 V ramp applied to the PZT and an injection current ramp from 124 mA to 40 mA. The experimentally optimized value of $\Delta I_{LD} / \Delta V_{PZT}$ is generally within 10% of the estimated value.

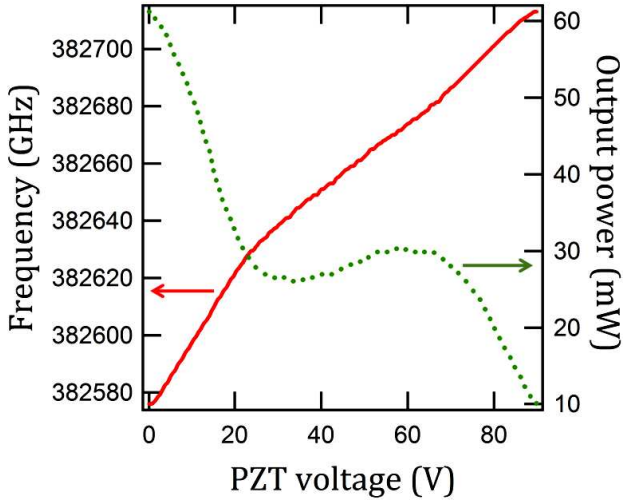

Fig. 3. Mode-hop-free scan of 137 GHz (from 382576 to 382713 GHz). The solid (red) curve shows the frequency (axis on the left) while the dotted (green) curve shows the output power (axis on the right), both plotted against the PZT voltage. The nonlinear dependence of the frequency and the output power on the PZT voltage may be due to the nonlinear dependence of the PZT displacement on $V_{PZT}$, coupled cavity effects of the internal and external cavities, and the variation of α and β over the long frequency scan.

In Fig. 4 we show the laser power transmitted through a Rubidium cell (containing both [85]Rb and [87]Rb) and through an Iodine cell (heated to 300˚C) during another MHF scan of 135 GHz. The scan is obtained with $\Delta I_{LD} / \Delta V_{PZT}$ = − 0.91 mA/V. The four $D_2$ absorption lines of Rubidium, two from [85]Rb (5s $^2S_{1/2}$, F=3 → 5p $^2P_{3/2}$ and 5s $^2S_{1/2}$, F=2 → 5p $^2P_{3/2}$) and two from [87]Rb (5s $^2S_{1/2}$, F=2 → 5p $^2P_{3/2}$ and 5s $^2S_{1/2}$, F=1 → 5p $^2P_{3/2}$), are clearly visible. The numerous absorption features in the Iodine transmission spectrum arise from different rovibrational transitions of $I_2$. We have verified that the frequencies of each of these Rb and $I_2$ absorption lines agree to within 1 GHz of the published line positions [18,19]. We also obtained a 129 GHz MHF tuning range around 781.6 nm (Fig. 5) using the same value of $\Delta I_{LD} / \Delta V_{PZT}$. The frequencies of these iodine absorption lines also match those found in Ref [19]. This is encouraging because it should be possible to patch together series of 100+ GHz scans by manually rotating the grating without readjusting the current ramp. Figs. 4 and 5 demonstrate the great potential of these ECDLs in spectroscopy where long single-mode frequency scans are desirable.

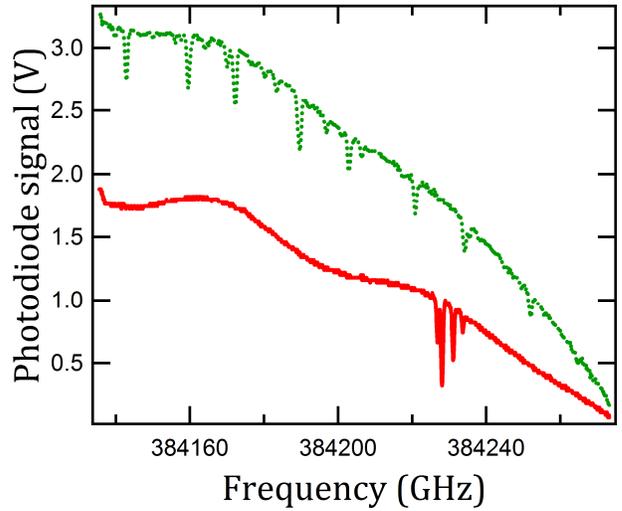

Fig. 4. Mode-hop-free scan of 135 GHz around 780.2 nm. Solid line (red): Laser transmission through a Rubidium vapor cell for a scan from 384137 to 384272 GHz. Dotted line (green): Transmission through an Iodine cell (heated to 300˚C).

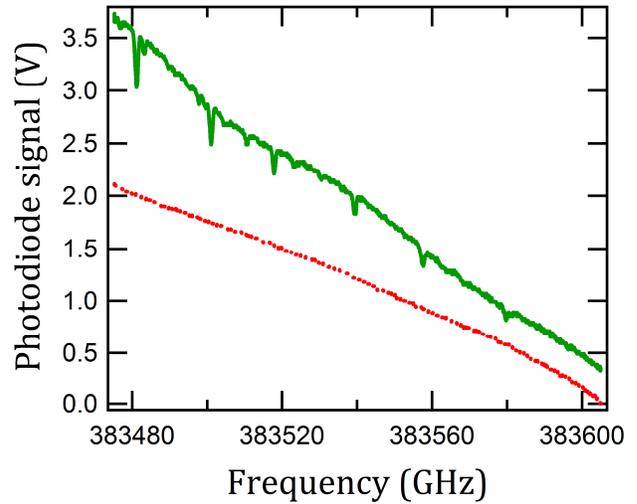

Fig. 5. Mode-hop-free scan of 129 GHz over a different frequency range (around 781.6 nm), with the same value of $\Delta I_{LD} / \Delta V_{PZT}$ as used for Fig. 4. Solid line (green): Laser transmission through an Iodine cell (heated to 300˚C). Dotted line (red): Transmission through a Rubidium vapor cell.

Finally, to understand which factors are more important than others for MHF tuning, we made another ECDL for which the pivot point is closer to the optimum location but observed no improvement in the tuning range. We then increased the external cavity length to $L_e$ = 18 mm. This change also implies that the pivot point was farther from the optimal position. Although the tuning range decreased, we were able to achieve over 100 GHz of MHF tuning. We attribute the decrease in MHF tuning to the increase in the length $L_e$ of the external cavity, and the corresponding decrease in

FSR. As discussed in the previous section, the relatively wide spectral width (due to finite angular width $\Delta\theta_g$) of the grating feedback makes the position of the pivot point less critical. The finite spectral width of the grating feedback was not considered in the analysis in Refs. 7-9, which predict that sub-millimeter accuracy is required for 100 GHz MHF scan.

## 5. CONCLUSION AND OUTLOOK

In conclusion, we have demonstrated a simple method to obtain large MHF tuning of a laser diode without AR-coating. We were able to achieve 137 GHz mode-hop-free tuning around 783.5 nm. This, to the best of our knowledge, is the widest MHF tuning of a simple cavity diode with or without an AR-coating. We found that, during the scan, it is extremely important that the frequency of the internal mode (tuned by the injection current) scans synchronously with the mode of the external cavity (tuned by a PZT that controls the external cavity length). It is easier to satisfy this condition for a shorter external cavity because a large FSR reduces the possibility of competing external cavity modes. The smallest external cavity we tried is around 15 mm; a smaller cavity is expected to result in an even higher MHF tuning. However, the tuning range may be limited by the injection current: we had to tune the current almost across the entire accessible range to obtain the 137 GHz scan. The tuning range is expected to increase for AR-coated diodes (where the internal cavity tends to be less important [20]). In addition a PZT with higher displacement is expected to increase the tuning range.

## 6. ACKNOWLEDGMENTS

This work is partially supported by the National Science Foundation (grant number CCF0829918). SD acknowledges Adeel Altaf and Dionysios Antypas for helpful discussions.